\documentclass[12pt]{article}

\begin{document}
%\renewcommand{\abstractname}{\ }

%УДК 517.958: 530.145
\title{Quantum Mechanics as Asymptotics of Solutions of Generalized Kramers Equation}
\author{E. M. Beniaminov}
\date{}
\maketitle
\begin{abstract}
We consider the process of diffusion scattering of a wave function given on the phase space.
In this process the heat diffusion is considered only along momenta. We write down the modified Kramers
equation describing this situation. In this model, the usual quantum description arises as
asymptotics of this process for large values of resistance of the medium per unit of mass of particle.
It is shown that in this case the process passes several stages. During the first short stage, the wave function
goes to one of ``stationary'' values. At the second long stage, the wave function varies in the subspace of
``stationary'' states according to the Schrodinger equation. Further, dissipation of the process leads
to decoherence, and any superposition of states goes to one of eigenstates of the Hamilton operator.
At the last stage, the mixed state of heat equilibrium (the Gibbs state) arises due to
the heat influence of the medium and the random transitions among the eigenstates of the Hamilton operator.

Besides that, it is shown that, on the contrary, if the resistance of the medium per unit of mass of particle
is small, then in the considered model, the density of distribution of probability $\rho =|\varphi |^2$
satisfies the standard Liouville equation, as in classical statistical mechanics.
\end{abstract}

\section{Description and some properties of the\\
model}
As in \cite{ben_expand}, we consider certain mathematical model of a process whose state at each moment of time is given
by a wave function, which is a complex valued function $\varphi (x, p)$, where $(x,p) \in R^{2n}$, on the phase space,
and $n$ is the dimension of the configuration space.
In contrast to quantum mechanics, where the wave function depends only on coordinates or only on momenta,
in our case the wave function depends both on coordinates and on momenta.
By analogy with quantum mechanics, it is assumed that wave functions obey the superposition principle, and the
density of probability $\rho_D (x,p)$ in a bounded domain of the phase space $(x, p) \in D \subset R^{2n}$, corresponding
to the wave function $\varphi(x, p),$ is given by the standard formula
\begin{equation}\label{rho}
\rho_D (x,p)=|\varphi(x, p)|^2 / \int_D |\varphi(x, p)|^2 dxdp.
\end{equation}

In quantum mechanics, the time evolution of the wave function can be defined by the Feynman path integral \cite{feinman}.
The Feynman principle assumes that if at the initial moment $t=t_0$
a wave function $\varphi (x_0,p_0, t_0)$ is given, then the value of the wave function at the point $(x,p)$ at the
moment $t=t_1$ is defined by the integral over all paths $\{x(t),p(t)\}$ joining the points $(x_0,p_0,t_0)$
and $(x,p,t_1)$, of the quantity $\exp\left(-\frac{i}{\hbar}\int _{t_0}^{t_1}[V(x(t))-{p^2(t)}/(2m)]dt\right)$, where
$\hbar$ is the Planck constant,
with respect to certain ``measure'' on paths defined by Feynman.

In contrast to Feynman's assumption, in the present paper we study the model in which the Feynman measure on paths
is replaced by the probability measure of the diffusion process (the heat Brownian motion) given by
the Kramers equation \cite{kramers}, \cite{vankampen}:
\begin{equation}\label{eq_kramers}
\frac{\partial f}{\partial{t}}=\sum_{j=1}^{n}
\biggl(
\frac{\partial V}{\partial x_j} \frac{\partial f}{\partial p_j}-
 \frac {p_j}{m} \frac{\partial f}{\partial x_j}
\biggr)
+\gamma \sum_{j=1}^{n}\frac{\partial }{\partial{p_j}}\biggl (p_j f
+kTm\frac{\partial{f} }{\partial{p_j}}\biggr) ,
\end{equation}
where
$f(x, p,t)$ is the density of probability distribution of the particle in the phase space at the moment of time $t$;
$m$ is the mass of the particle;
$V(x)$ is the potential function of the external forces acting on the particle;  $\gamma=\beta /m$ is the resistance
coefficient of the medium per unit of mass; $k$ is the Boltzmann constant;
$T$ is the temperature of the medium.

This is the classical Kramers equation describing the diffusion motion of a particle in the phase space
under action of external forces defined by the potential function $V(x)$, the heat medium with temperature
$T$, and the medium resistance per unit of mass $\gamma.$

Consider the following modified Kramers equation for the wave function $\varphi(x, p,t)$:
\begin{equation}\label{eq_diff}
\frac{\partial\varphi}{\partial{t}}=A\varphi
+\gamma B{\varphi},
\end{equation}
\begin{equation}\label{def_A}
\mbox{where }\ \ \ \ \ \ A\varphi =\sum_{j=1}^{n}
\biggl(
\frac{\partial V}{\partial x_j} \frac{\partial\varphi}{\partial p_j}-
 \frac {p_j}{m} \frac{\partial \varphi}{\partial x_j}
\biggr)
-\frac{i}{\hbar}
\biggl(mc^2+V-\sum_{j=1}^{n}\frac{p^2_j}{2m} \biggr)\varphi \ \ \ \ \ \ \
\end{equation}
\begin{equation}\label{def_B}
\mbox{и }\ \ \ \ \  B{\varphi}=
\sum_{j=1}^{n}\frac{\partial}{\partial{p_j}}\left( \biggl( p_j +i\hbar \frac{\partial}{\partial{x_j}}\biggr){\varphi}
+kTm\frac{\partial{\varphi} }{\partial{p_j}}\right). \ \ \ \ \ \ \ \nonumber
\end{equation}
Equation (\ref{eq_diff}) is obtained from the Kramers equation (\ref{eq_kramers}) by adding to the right hand side
the summand of the form
$-{i}/{\hbar} (mc^2+V-{p^2}/(2m) )\varphi$ and by replacing multiplication of the function $\varphi $ by $p_j$
in the diffusion operator by the action of the operator
$(p_j+i\hbar \partial/{\partial x_j})$ on the function $\varphi $.

Adding the summand $-{i}/{\hbar} (mc^2+V-{p^2}/(2m) )\varphi$ is related with the additional physical requirement
that the wave function at the point $(x, p)$ oscillates harmonically with the frequency
${1}/{\hbar} (m c^2+V-{p^2}/(2m) )$ in time.
%(или, что то же самое, колеблется с  частотой $mc^2/\hbar$
%в собственном времени в системе координат, связанной с частицей массы $m$, движущейся в точке $x$ с импульсом $p$).

The requirement of harmonic oscillation of the wave function $\varphi $ at the point $(x, p)$ with large
frequency
 ${1}/{\hbar} (m c^2+V-{p^2}/(2m) )$, in the case when $ m c^2$ is much greater than $V$, gives that the shift of the
wave function with respect to coordinate $x_j$ with conservation of the proper time at the point $(x, p)$
yields the phase shift
in the oscillation of the function $\varphi$. This also gives that the operator of infinitesimal shift
$\partial/{\partial x_j} $ is replaced by the operator
$\partial/{\partial x_j}-ip_j/\hbar$. (For a more detailed explanation, see \cite{ben_expand}.) Respectively,
if one multiplies this operator by $i\hbar$, then one obtains the operator $p_j+ i\hbar \partial/{\partial x_j}$
used in the modified diffusion operator $B$.

Let us proceed to the study of equation (\ref{eq_diff}).

In order to separate mathematics from physics in this equation, let us make a change of variables and let us pass to
dimensionless quantities:
\begin{equation}\label{new_var}
t'=\gamma t,\ \ \ \ p'=\frac{p}{ \sqrt {kTm}}, \ \ \ \ x'=\frac{ \sqrt {kTm}}{\hbar}x,\ \ \ \ V'(x)=\frac{V(x)}{kT}.
\end{equation}

In the new variables equation (\ref{eq_diff}) takes the following form:
\begin{equation}\label{eq_diff'}
\frac{\partial\varphi}{\partial{t'}}=\frac{kT}{\gamma \hbar}A'\varphi+ B'{\varphi},
\end{equation}
\begin{equation}\label{def_A'}
\mbox{where }\ \ \ \ \ \ A'\varphi =\sum_{j=1}^{n}
\biggl(
\frac{\partial V'}{\partial x'_j} \frac{\partial\varphi}{\partial p'_j}-
  {p'_j} \frac{\partial \varphi}{\partial x'_j}
\biggr)
-{i}
\biggl(\frac{mc^2}{kT}+V'-\sum_{j=1}^{n}\frac{(p'_j)^2}{2} \biggr)\varphi \ \ \ \ \ \mbox{и }
\end{equation}
\begin{equation}\label{def_B'}
B'{\varphi}=
\sum_{j=1}^{n}\frac{\partial}{\partial{p'_j}}\left( \biggl( p'_j +i \frac{\partial}{\partial{x'_j}}\biggr){\varphi}
+\frac{\partial{\varphi} }{\partial{p'_j}}\right). \nonumber
\end{equation}

The parameter of the model described by equation~(\ref{eq_diff'}), is the dimensionless quantity $kT/(\gamma \hbar)$,
which we denote by $\varepsilon $.

Let us assume that $\varepsilon =kT/(\gamma \hbar) $ is a small quantity which is the small perturbation parameter
in equation~(\ref{eq_diff'}) with the non-perturbed equation ${\partial\varphi}/{\partial{t'}}= B'\varphi  $,
i.~e., the equation
\begin{equation}\label{eq_B}
\frac{\partial\varphi}{\partial{t'}}=
\sum_{j=1}^{n}\frac{\partial}{\partial{p'_j}}\left( \biggl( p'_j +i \frac{\partial}{\partial{x'_j}}\biggr){\varphi}
+\frac{\partial{\varphi} }{\partial{p'_j}}\right).
\end{equation}
Note that the smallness of the quantity $\varepsilon =kT/(\gamma \hbar) $  requires that the friction
coefficient of the medium per unit of mass $\gamma=\beta /m=(k/ \hbar) T/\varepsilon = 1.3\cdot 10^{11}T/\varepsilon $
be greater than $1.3 \cdot10^{11}T$,
since $k/ \hbar=1.3\cdot 10^{11}$.

Let us substitute into equation~(\ref{eq_B}) the Fourier integral presentation of $\varphi(x',p',t')$
with respect to $x'$:
\begin{equation} \label{fur}
  \varphi(x',p',t') =\frac{1}{(2\pi)^{n/2}}
\int_{R^n}\tilde\varphi(s',p',t')e^{i s' x'}ds',
\end{equation}
\begin{equation} \label{fur_1}
\mbox{where   }\ \
  \tilde\varphi(s',p',t') =\frac{1}{(2\pi)^{n/2}}
\int_{R^n}\varphi(x',p',t')e^{-{i s' x' }}dx'.
\end{equation}

We obtain that $ \tilde\varphi(s',p',t')$ satisfies the equation
\begin{equation}\label{eq_L_tilde}
\frac{\partial\tilde\varphi}{\partial{t'}}=
\sum_{j=1}^{n}\frac{\partial}{\partial{p'_j}} \biggl(( p'_j -s'_j){\tilde\varphi}
+\frac{\partial{\tilde\varphi} }{\partial{p'_j}}\biggr).
\end{equation}

The operator of the right hand side of this equation is well known (see, for example, \cite{kamke}).
This operator has a full set of eigenfunctions in the space of
functions tending to zero as $|p'|$ tends to infinity.
The eigenvalues of this operator are nonpositive integers.
The eigenfunctions corresponding to the eigenvalue 0 have the form
$$\tilde\varphi_0(s', p')= \frac{1}{(2\pi)^{n/2}}\tilde\psi(s') e^{-\frac{(p'-s')^2}{2}},$$
where $\tilde\psi(s')$ is an arbitrary complex valued function of $s'\in R^n$ .

The rest of eigenfunctions are obtained as derivatives of the functions $\tilde\varphi_0(s', p')$ with respect to
$p'$, and have eigenvalues $-1, -2,  ...$, respectively, depending on the degree of derivative, and the
projector $P_0$ to the subspace of eigenfunctions with eigenvalue 0 has the form
\begin{equation}\label{P_0_tilde}
 \tilde\varphi_0(s', p')=P_0 \tilde\varphi=\frac{1}{(2\pi)^{n/2}}\tilde\psi(s') e^{-\frac{(p'-s')^2}{2}},\ \
\mbox{where} \ \ \tilde\psi(s')=\int_{R^n}\tilde\varphi (s',p')dp'.
\end{equation}
Hence, considering equation~(\ref{eq_L_tilde}) in the basis of these eigenfunctions, we obtain that each
solution $ \tilde\varphi(s',p',t')$ of this equation tends exponentially in time with exponent $-1 $ to a
stationary solution of the form $\tilde\varphi_0$.  Therefore, taking into account
the presentation~(\ref{fur}) of the function $\varphi(x',p',t)$ via $ \tilde\varphi(s',p',t')$,
we obtain that ``stationary''
 solutions $\varphi_0(x',p')$ of equation~(\ref{eq_diff'}) look as follows:
$$
\varphi_0(x',p')=\frac{1}{(2\pi)^{n}}
\int_{R^n}\tilde \psi(s') e^{-\frac{(p'-s')^2}{2}} e^{i s' x'}ds'.
$$

Let us present the function $\tilde\psi(s')$, in its turn, as the Fourier integral:
$$
  \tilde\psi (s') =\frac{1}{(2\pi)^{n/2}}
\int_{R^n}\psi (y')e^{-{i s' y'}}dy'.
$$
Substituting this presentation into the preceding expression and integrating over $s'$, we obtain
$$
\varphi_0(x',p')=\frac{1}{(2\pi)^{3n/2}}
\int_{R^{2n}}\psi (y') e^{-\frac{(p'-s')^2}{2}}
e^{i s'(x'- y') }ds'\ dy'
$$
$$
=\frac{1}{(2\pi)^{n}}
\int_{R^n}\psi (y') e^{-\frac{(x'-y')^2}{2}}
e^{i p'(x'- y') } dy'
$$
or, taking into account~(\ref{P_0_tilde}),
\begin{equation}\label{P'_0}
\varphi_0=P_0 \varphi =\frac{1}{(2\pi)^{n}}
\!\!\int\limits_{R^n}\!\!\psi (y') e^{-\frac{(x'-y')^2}{2}}
e^{i p'(x'- y') } dy',\  \mbox{where}\ \psi (y') =\!\!\int\limits_{R^n}\varphi (y',p') dp'.
\end{equation}
Thus, if $\varepsilon $  is small, then at the time $t'$ of order $1$, a solution of equation~(\ref{eq_diff'}),
starting from an arbitrary function $\varphi,$
will become close to a function of the form $\varphi_0$ which, in the initial coordinates~(\ref{new_var}), reads:
$$\varphi_0(x,p)=\frac{1}{(2\pi\hbar)^{n}}
\int_{R^n}\psi (y) \exp\left({-\frac{kTm(x-y)^2}{2\hbar^2}}\right )
\exp\left(\frac{i p(x- y) }{\hbar}\right) dy.
$$
 Later the solution of equation (\ref{eq_diff}), evolving in time, will be close to the subspace of ``stationary''
 functions of the form $\varphi_0$.

Note that the ``stationary'' functions obtained here coincide, up to a normalizing factor, with the
``stationary'' functions of Theorem~1 obtained in the work \cite{ben_expand}, if one puts $kTm/\hbar$ equal to $b/a$,
where $a^2$ and $b^2$  are the diffusion coefficients with respect to coordinates and momenta in the model
considered in \cite{ben_expand}. Hence some results obtained in \cite{ben_dan}, \cite{ben_expand}, hold true
also in our case. We shall cite them without proof.

The results are obtained by the perturbation theory method up to second order, of equation~(\ref{eq_diff'})
${\partial \varphi }/{\partial t}=\varepsilon A'\varphi +B'\varphi$ for $\varepsilon  \ll 1$,
where $A'$ is a skew Hermitian operator and $B'$ is an operator with nonpositive discrete spectrum.

Note that this equation does not preserve the norm of the function $\varphi $, which we normalize to obtain
the distribution $\rho (x,p)$.
The corresponding equation preserving the norm, which we actually study, reads
${\partial \varphi }/{\partial t}=\varepsilon A'\varphi +B'\varphi +k\varphi,$
where $k=- (\langle B'\varphi ,\varphi \rangle+\langle \varphi ,B'\varphi \rangle)/(2\langle \varphi ,\varphi \rangle)$,
but it is nonlinear.

\section {The main results}
Denote by $H(x, p)=p^2/(2m)+V(x) +mc^2$ the Hamilton function of the system.

{\it {\bf Theorem 1}.
The motion described by equation~(\ref{eq_diff}), for small $\varepsilon =kT/(\gamma\hbar)$
asymptotically splits into rapid and slow motion.

1)  After rapid motion the arbitrary wave function $\varphi(x, p, 0)$
goes, in time of order $1/\gamma $, to a function which after normalization has the form
\begin{eqnarray}\label{view_varphi'}
\varphi_0(x, p)=\frac{1}{(2\pi{\hbar})^{n/2}}
\!\int\limits_{R^n}\!\!\psi(y)\chi(x-y)
e^{{{i  p(x-y)}/{\hbar}}}
dy,\\
\label{chi_def'}\mbox{where          }\ \ \ \ \ ||\psi ||=1 \ \ \ \ \mbox{and} \ \ \ \
  \chi(x-y)=\left(\frac{kTm}{\pi\hbar^2}\right)^{n/4}e^{-{{kTm}(x-y)^2}/{(2\hbar^2)}},
\end{eqnarray}
  The wave functions of the form~(\ref{view_varphi'}) form a linear space.
The elements of this subspace are parameterized by the wave functions $\psi(y)$ depending only on coordinates $y\in R^n$.

2) The slow motion starting from the wave function $\varphi_0(x, p)$ of the form~(\ref{view_varphi'}) with nonzero
function $\psi(y)$, goes inside the subspace and is parameterized by the wave function $\psi(y,t)$
depending on time. The function $\psi(y,t)$ satisfies the Schrodinger equation of the form
$ i\hbar {\partial \psi}/{\partial t} = \hat{H}\psi$,  where
\begin{equation}\label{hatH}
{\hat H}\psi =- \frac{\hbar^2}{2m}\biggl(\sum_{k=1}^{n}\frac{\partial^2 \psi }{\partial{y^2_k}}\biggr)
+V(y) \psi+\frac{kT}{2}n\psi+mc^2\psi.
\end{equation}}

Proof of the first part of Theorem~1 is given in \cite{ben_expand}. Proof of the second part of the Theorem
is given in Appendix~1 to the present paper.

{\it {\bf Theorem~2}.
The projection operator $P_0$ transforming an arbitrary integrable function $\varphi(x,p)$
on the phase space into the function of the form
(\ref{view_varphi'}) (but without normalization), obtained after rapid motion described in Theorem~1, reads as follows:
 \begin{eqnarray}\label{p_0}
P_0\varphi =
\frac{1}{(2\pi{\hbar})^{n}}
\!\int\limits_{R^{n}}\!\!\ \psi (y)e^{-\frac{kT(x-x')^2}{2\hbar^2}}e^{\frac{i  p(x-x')}{\hbar}} dy,
 \ \ \mbox{where}\ \ \psi (y) =\int\limits_{R}\!\!\ \varphi(y,p) dp.
\end{eqnarray}
}

This Theorem follows from the definition of the operator $P_0$ and from formula~(\ref{P'_0}) expressed in the
initial coordinates~(\ref{new_var}).

{\it {\bf Theorem~3}.
If $\psi(x)$ is a wave function on the configuration space
and $\varphi_0(x,p)$ is the wave function on the phase space corresponding to it by formula~(\ref{view_varphi'}),
then the density of probability $\rho(x,p)=|\varphi_0(x,p)|^2$ in the phase space is given by the following formula:
\begin{eqnarray}\label{rho_psi}
\rho (x,p)=
\frac{1}{(2\pi{\hbar})^{n}} \left(\frac{kTm}{4\pi\hbar^2}\right)^{n/2}
\int\limits_{R^{2n}} \! \psi\!\left(x+\frac{x''-x'}{2}\right)\psi^{*}\!\left(x+\frac{x''+x'}{2}\right)  \\ \nonumber
\exp\!\left({-\frac{kTm(x'')^2}{4\hbar^2}}\right)
\exp\!\left({-\frac{kTm(x')^2}{4\hbar^2}}\right)
\exp\!\left({\frac{i x' p}{\hbar}}\right)
dx'' dx'.
\end{eqnarray}
}

In contrast to quasidistributions
$$W(x,p)=\frac{1}{(2\pi{\hbar})^{n}}
\int\limits_{R^{n}}\!\!\psi\!\left(x-\frac{x'}{2}\right)\psi^{*}\!\left(x+\frac{x'}{2}\right)
\exp\!\left({\frac{i x' p}{\hbar}}\right)dx'
$$
defined by Wigner~\cite{wigner},
the density $\rho(x,p)$ in the phase space, given by the expression~(\ref{rho_psi}), is always nonnegative.
Its expression differs from the expression of the Wigner function by exponents under the integral, which
yield smoothing with respect to distribution densities close to the delta-functions.

Proof of Theorem~3 is given in \cite{ben_expand}.

The algebra of observables given by real functions on the phase space,
averaged by densities of probability distributions of the form~(\ref{rho_psi}), has been studied in~\cite{ben_arxiv}.

Since the probability distribution $\rho(x)$ in the configuration space is expressed by the formula
$\rho(x)=\int_{R^3}\rho(x,p) dp,$ by integrating the expression~(\ref{rho_psi}) over $p$
 we obtain the following statement.

{\it  {\bf Corollary 1}.
If $\psi(x)$ is the wave function on the configuration space, then
the corresponding density $\rho(x)$ in the configuration space
is given by the formula
\begin{equation}\label{rho_x}
\rho (x)=\int_{R^3}|\psi(y)|^2 \chi^2(x-y) dy,
\end{equation}
where $\chi(x,y)$ is expressed by relation~(\ref{chi_def'}).
That is, $\rho (x)$ is obtained from $|\psi(x)|^2$ by smoothing (convolution)
with respect to the density of the normal distribution with dispersion $\hbar^2/(2kTm)$,
and the exactness of defining coordinate is bounded by the quantity $\sim \hbar/ \sqrt {2kTm}$ called the
de Broglie length of the heat wave.
}

{\it {\bf Theorem~4}. If the number $\varepsilon =kT/(\gamma \hbar) \ll 1$, then the first order
perturbations of the zero eigenvalue of the operator $B'$ in equation~(\ref{eq_diff'}) equal to the eigenvalues
of the operator $-i/(\gamma \hbar)\hat H$, where
$\hat H$ is the Hamilton operator given by formula~(\ref{hatH}). If the real parts of the
second order perturbations corresponding to these first order perturbations are different from one another,
then any solution of equation~(\ref{eq_diff}) will go in time proportional to
$1/(\gamma \varepsilon^2)=\gamma \hbar^2/(kT)^2$, to one of the eigenstates of the Hamilton operator.
}

Proof of the Theorem is given in Appendix~2.

This theorem describes the process called in the literature by decoherence of quantum states \cite{zeh},
\cite{zurek}, \cite{mensky}. The form of the estimate of the time of decoherence given in Theorem~4, somewhat differs
from the form of the estimate given in the literature, for instance in \cite{zurek}.
The study of the correspondence of these estimates is the subject of another future work.

According to Theorem~4 the process described by equation~(\ref{eq_diff}), in time proportional to
$\gamma \hbar^2/(kT)^2$, goes to one of eigenstates of the Hamilton (energy) operator.
Further, on large scales of time the system, under the action of the heat medium, will jump from one eigenstates
to others due to large deviations of the random process. In the limit as $t\rightarrow \infty$
the system will go to the mixed state corresponding to the heat equilibrium Gibbs state.

We have studied equation~(\ref{eq_diff}) for small value of $\varepsilon =kT/(\gamma \hbar)$ and obtained that
the process described by this equation is asymptotically close to the process having the standard quantum description.
Let us now consider the same equation in the case when the quantity $\varepsilon $ is large.

{\it {\bf Theorem~5}. If $\varepsilon =kT/(\gamma \hbar) \gg 1$, i.~e. $\gamma \hbar/(kT) \ll 1$, where $\gamma =\beta/m$
is the resistance of medium per unit of mass, then the operator $B$ in equation~(\ref{eq_diff}) can be neglected,
and in this case the density of probability distribution $\rho (x,p,t)=\varphi(x, p, t)\varphi^*(x, p, t)$
satisfies the following classical Liouville equation:
\begin{equation}\label{liuvill}
\frac{\partial \rho }{\partial{t}}=\sum_{j=1}^{n}
\biggl(
\frac{\partial V}{\partial x_j} \frac{\partial \rho}{\partial p_j}-
 \frac {p_j}{m} \frac{\partial \rho}{\partial x_j}
\biggr).
\end{equation}
}

Proof of the Theorem follows from the fact that equation~(\ref{eq_diff}) without the operator $B$
is a partial differential equation of first order, consisting of sum of the Liouville operator and the
operator of multiplication by a function.
The solution $\varphi (x,p,t)$ of such equation is obtained from the initial state
$\varphi (x,p)=\varphi (x,p,0)$ and the characteristics $ x(t), p(t)$  of the equation with the following
initial conditions: $ x(0)=x, \ p(0)=p$, in the following form:
$$\varphi (x,p,t)=\varphi (x(t), p(t)) \exp \left(-i /\hbar\int_0^t (H(x(t), p(t))-p^2(t)/(2m))dt \right).$$
Respectively, we have:
$$\rho (x,p,t)=\varphi(x, p, t)\varphi^*(x, p, t)=\varphi (x(t), p(t))\varphi^* (x(t), p(t))=\rho(x(t), p(t)). $$
Therefore, the phase of the wave is inessential, and the density $\rho (x,p,t)$ satisfies the Liouville equation.

Thus, Theorem~5 states that for small value of $\beta \hbar/(kTm)$ the process
described by equation (\ref{eq_diff}) is asymptotically close to the process with the classical (non-quantum)
description of the motion of the particle. This case arises when the mass $m$ of the particle is large
relative to the medium resistance $\beta $.

\section{Conclusion}
In quantum optics, one now widely uses methods of study of dynamics of quantum processes in the phase space,
see, for example, \cite{Shlyajx}, where, in particular, one considers dynamics of wave packets and interference
in the phase space. On this way, one manages to model dynamics of ions in traps,
optics of atoms in quantum light fields, etc. ``This approach which stresses the fundamental role of phase
variables allows one to expose and interprete very clearly various branches of quantum optics...''
(from the abstract of the book  \cite{Shlyajx}).

These achievements lead to an assumption that it would be useful to consider not only the behavior of
distributions in the phase space, as it is done, for example, in \cite{Shlyajx}, \cite{manko}, \cite{khrennikov},
but also to introduce the wave function itself on the phase space.

In the present work, following this direction, we construct a diffusion equation for the wave function with large
frequency of oscillations in the phase space, describing the process of heat scattering of a wave in the phase space.
In the presented model one meets both classical and quantum mechanics behavior of the particle.
If the quantity $\varepsilon =kTm/(\beta \hbar) $ is small,
then in this model the behavior of the particle, after short transition stage
(of order $m/\beta < \hbar/kT=0.77 \cdot 10^{-12}/T$ sec.)
amounts to the standard picture of quantum mechanics with the Heisenberg indeterminacy principle and with the
Schrodinger equation describing the dynamics. And if the quantity $\varepsilon $ is large, then the particle
behaves according to classical mechanics, and the density of probability distribution of the particle in the
phase space satisfies the classical Liouville equation.

In the general case, the behavior of the particle described by this model is of mixed nature.
It would be interesting to analyze the model in this case and compare it with results of experiments
for particles with intermediate values of $\varepsilon $.

One should also compare with experimental data the estimate of time of decoherence given in Theorem~4. Besides that,
one should find a theoretical estimate of the transition time of the quantum system to the mixed state of heat
equilibrium in this model.

One should also acknowledge numerous works of predecessors, due to which the subject of this paper arose,
the problem has been stated and the methods of its solution have been found. This is a separate large work.
In science, posing right questions gives no less than the results obtained. Bright examples of it are the questions
of A.~Einstein, due to which quantum mechanics was founded and develops up to now.

In the works of V.~P.~Maslov \cite{maslov1, maslov2} one has already studied certain problem of
description of motion of a distribution of charges in the phase space under action of a random field. In
\cite{maslov1} under certain assumptions it is proven that if the initial distribution of charges,
depending on coordinates and momenta, belongs to certain subspace, parameterized by complex valued functions
depending on coordinates, then the distribution does not leave this subspace, and the dynamics of such system
is described by the corresponding Schrodinger equation. This result has certainly influenced the author
while posing the problem of the present work.
\bigskip

{\bf Acknowledgements:} I am deeply grateful in my heart to Professor G.~L.~Litvinov for the many year attention
to my work, for understanding, and for help in the right organization and exposition of the material. 
I am grateful to  professor A.~V.~Stoyanovsky, who translated this paper to English.

\newpage
\section*{Appendices}

%\begin{flushright}
%{\large Приложение 1}
%\end{flushright}

\subsection*{Appendix 1. Proof of Part 2 of Theorem 1}
Consider equation~(\ref{eq_diff}) in the dimensionless system of variables~(\ref{new_var}). In these variables,
the equation takes the form~(\ref{eq_diff'}), and ``stationary'' solutions,
to which arbitrary solutions of equation~(\ref{eq_diff'}) tend at time $t'$ of order 1, have the form~(\ref{P'_0}).

Let $\varphi_0(x',p',t')$ be a function of the form~(\ref{P'_0}) corresponding to the function $\psi(y',t')$.
Let us substitute this expression into equation~(\ref{eq_diff'}), and let us take projection of both parts
of this equation to the space of functions $\psi(y',t')$ by formula~(\ref{P'_0}). We have:
$$
\int\limits_{R^n}\frac{\partial \varphi_0}{\partial t'}dp' =
\int\limits_{R^n}\left(\frac{kT}{\gamma \hbar }A'+B' \right)\varphi_0dp'
$$
or, taking into account that $B'\varphi _0=0$,  after substitution of expression $\varphi_0(x',p',t')$ in the
form~(\ref{P'_0}), we obtain:
$$
\frac{1}{(2\pi)^{n}}\!\!\int\limits_{R^{2n}}\!\!\frac{\partial \psi (y',t')}{\partial t'} e^{-\frac{(x'-y')^2}{2}}
e^{i p'(x'- y') } dy'dp'\ \ \ \ \ \ \ \ \ \
$$
$$\ \ \ \ \ \ \ \ \ \ \ \ \ \ \ \ \ \ \ \ \ =\frac{kT}{\gamma \hbar }\frac{1}{(2\pi)^{n}}
\!\!\int\limits_{R^{2n}}\!\!A'\psi (y',t') e^{-\frac{(x'-y')^2}{2}}
e^{i p'(x'- y') } dy' dp'.
$$
Let us integrate the right hand side of this equality over $p'$ and over $y'$. Noting that in the right hand side
of the equality we have the delta function, we obtain:
$$
\frac{\partial \psi (x',t')}{\partial t'}=\frac{kT}{\gamma \hbar }\frac{1}{(2\pi)^{n}}
\!\!\int\limits_{R^{2n}}\!\!A'\psi (y',t') e^{-\frac{(x'-y')^2}{2}}
e^{i p'(x'- y') } dy' dp'.
$$
Taking into account expression~(\ref{def_A'}) for operator $A'$, we deduce from the latter equality that
\begin{eqnarray}\label{H'full}
\frac{\partial \psi }{\partial t'}\!\!\!\!&=&\!\!\!\!\frac{kT}{\gamma \hbar }\frac{1}{(2\pi)^{n}}
\!\!\int\limits_{R^{2n}}\!\!\!\!\left(\sum_{j=1}^{n}
\biggl(
\frac{\partial V'}{\partial x'_j} \frac{\partial}{\partial p'_j}-
  {p'_j} \frac{\partial }{\partial x'_j}
\biggr)
-{i}
\biggl(\frac{mc^2}{kT}+V'-\sum_{j=1}^{n}\frac{(p'_j)^2}{2} \biggr)\!\!\right)\nonumber\\
&&\!\!\!\!\times\psi (y',t') e^{-\frac{(x'-y')^2}{2}}
e^{i p'(x'- y') } dy' dp'=
\frac{kT}{\gamma \hbar }(I_1+I_2+I_3+I_4),
\end{eqnarray}
where
\begin{eqnarray}\label{def_I_1}
I_1&=&\frac{1}{(2\pi)^{n}}\!\!\int\limits_{R^{2n}}\sum_{j=1}^{n}\frac{\partial V'(x')}{\partial x'_j} \frac{\partial}{\partial p'_j}
\left(\psi (y',t') e^{-\frac{(x'-y')^2}{2}} e^{i p'(x'- y') } \right) dy' dp';\\
I_2&=&-\frac{1}{(2\pi)^{n}}\!\!\int\limits_{R^{2n}}\sum_{j=1}^{n}{p'_j} \frac{\partial }{\partial x'_j}\left(
\psi (y',t') e^{-\frac{(x'-y')^2}{2}}
e^{i p'(x'- y') }  \right)dy' dp';\label{def_I_2}\\
I_3&=&-i\frac{1}{(2\pi)^{n}}\!\!\int\limits_{R^{2n}}\biggl(\frac{mc^2}{kT}+V'(x')\biggr)
\psi (y',t') e^{-\frac{(x'-y')^2}{2}}
e^{i p'(x'- y') } dy' dp';\label{def_I_3}\\
I_4&=&i\frac{1}{(2\pi)^{n}}\!\!\int\limits_{R^{2n}}\sum_{j=1}^{n}\frac{(p'_j)^2}{2}
\psi (y',t') e^{-\frac{(x'-y')^2}{2}}
e^{i p'(x'- y') } dy' dp'.\label{def_I_4}
\end{eqnarray}

Consider the integral $I_1$ given by expression~(\ref{def_I_1}). Let us exchange summation and integration,
let us put behind the sign of integral expressions not depending on integration variables, let us compute
the derivatives with respect to $p'_j$,  and let us integrate the remaining integrals over $p'$ and $y'$.
We obtain:
\begin{eqnarray}\label{res_I_1}
I_1=\frac{i}{(2\pi)^{n}}\sum_{j=1}^{n}\frac{\partial V'(x')}{\partial x'_j}\int\limits_{R^{2n}}
\psi (y',t') e^{-\frac{(x'-y')^2}{2}}(x'_j- y'_j) e^{i p'(x'- y') } dy' dp'=0
\end{eqnarray}

Consider the integral $I_2$ given by expression~(\ref{def_I_2}). Let us exchange summation and integration,
let us transfer the derivatives with respect to $x'_j$ behind the sign of integral, let us replace the
expressions $p'_j \exp(ip'(x'-y'))$ by equal expressions
$i\partial \ \exp(ip'(x'-y'))/(\partial y'_j)$, and let us integrate the obtained integrals by parts. We have:
\begin{eqnarray}\label{res_I_2}
I_2&=&-\frac{1}{(2\pi)^{n}}\sum_{j=1}^{n}\frac{\partial }{\partial x'_j}\int\limits_{R^{2n}}
\psi (y',t') e^{-\frac{(x'-y')^2}{2}}i\frac{\partial  }{\partial y'_j}e^{i p'(x'- y') } dy' dp'\nonumber\\
&=&\frac{i}{(2\pi)^{n}}\sum_{j=1}^{n}\frac{\partial }{\partial x'_j}\int\limits_{R^{2n}}
    \frac{\partial \psi (y',t')}{\partial y'_j}e^{-\frac{(x'-y')^2}{2}}e^{i p'(x'- y') } dy' dp'\nonumber\\
&&-\frac{i}{(2\pi)^{n}}\sum_{j=1}^{n}\frac{\partial }{\partial x'_j}\int\limits_{R^{2n}}
    \psi (y',t')(x'_j-y'_j)e^{-\frac{(x'-y')^2}{2}}e^{i p'(x'- y') } dy' dp'\nonumber\\
&=&i\sum_{j=1}^{n}\frac{\partial^2  \psi (x',t')}{\partial (x'_j)^2}.
\end{eqnarray}

Consider the integral $I_3$ given by expression~(\ref{def_I_3}). Let us transfer behind the sign of integral
the expressions not depending on the integration variables, and let us integrate the remaining integral over $p'$
and $y'$.
We obtain:
\begin{eqnarray}\label{res_I_3}
I_3&=&-i\biggl(\frac{mc^2}{kT}+V'(x')\biggr) \frac{1}{(2\pi)^{n}}\int\limits_{R^{2n}}
\psi (y',t') e^{-\frac{(x'-y')^2}{2}} e^{i p'(x'- y') } dy' dp'\nonumber\\
&=&-i\biggl(\frac{mc^2}{kT}+V'(x')\biggr)\psi (x',t').
\end{eqnarray}

Consider the integral $I_4$ given by expression~(\ref{def_I_4}). Let us exchange summation and integration,
let us transfer $1/2$ behind the sign of integral, let us replace the expression $(p'_j)^2 \exp(ip'(x'-y'))$
by the second derivative of the function
$-\exp(ip'(x'-y'))$ with respect to $y'_j$,
and let us integrate the obtained integrals by parts. We have:
\begin{eqnarray}\label{res_I_4}
I_4&=&-\frac{i}{2}\frac{1}{(2\pi)^{n}}\sum_{j=1}^{n}\int\limits_{R^{2n}}
\psi (y',t') e^{-\frac{(x'-y')^2}{2}}\frac{\partial^2  }{\partial (y'_j)^2}e^{i p'(x'- y') } dy' dp'\nonumber\\
&=&-\frac{i}{2}\frac{1}{(2\pi)^{n}}\sum_{j=1}^{n}\int\limits_{R^{2n}}
    \left(\frac{\partial^2 \psi }{\partial (y'_j)^2}-2\frac{\partial \psi }{\partial y'_j}(x'_j-y'_j)+
\psi (x'_j-y'_j)^2+\psi \right)\nonumber\\
&& \ \ \ \ \ \ \ \ \ \ \ \ \ \ \ \ \ \ \ \ \ \ \ \ \ \ \ \ \ \ \ \ \ \ \ \ \ \ \ \ \times
e^{-\frac{(x'-y')^2}{2}}e^{i p'(x'- y') } dy' dp'\nonumber\\
&=&-\frac{i}{2}\left(\sum_{j=1}^{n}\frac{\partial^2  \psi (x',t')}{\partial (x'_j)^2}+ n \psi (x',t')\right).
\end{eqnarray}

Let us substitute the obtained expressions for integrals $I_1,\ldots,I_4$ into equality~(\ref{H'full}), let us sum up
similar terms, and let us transfer $-i$ behind the brackets.
We obtain,
$$
\frac{\partial \psi }{\partial t'}= -\frac{i}{ \hbar }\frac{kT}{\gamma}
\left(
-\frac{1}{2}\frac{\partial^2}{\partial (x'_j)^2}+V'+ \frac{mc^2}{kT}+\frac{n}{2}
\right)\psi (x',t').
$$

If in the obtained equality we pass to the initial coordinate system~(\ref{new_var}), then we obtain the
equality~(\ref{hatH}) required in Part~2 of Theorem~1.

\newpage
\subsection*{Appendix 2.  Proof of Theorem 4}
Consider equation~(\ref{eq_diff'}) of the form $\partial \varphi /\partial t'=\varepsilon A'\varphi +B'\varphi $.
Let $P_0$ be the projector onto the subspace of eigenfunctions of operator $B'$ with eigenvalue 0.
Then, by definition of $P_0$, we have the equalities:
$$P_0 P_0=P_0 ;\ \ \ \ \ \ P_0 B'=B' P_0=0.$$
We shall denote by $\varphi_0$ a function belonging to the image of projector $P_0$, i.~e.
$\varphi_0=P_0\varphi_0.$

Note that by construction of operator $\hat H$ from Theorem~1, given in Appendix~1,
the operator $ \varepsilon P_0 A'$  on the subspace of functions $\varphi_0$, in the presentation by the functions
$\psi(y)$ has the form $-i/(\gamma \hbar)\hat H$.
Hence the eigenvalues of operators $\varepsilon P_0 A'$ and $-i/(\gamma \hbar)\hat H$ coincide.
Since the operator $\hat H$ is self-adjoint, it has a complete system of eigenfunctions, and therefore
the operator $\varepsilon P_0 A'$  on the subspace of values of the projector $P_0$ also has a complete system
of eigenfunctions.

Let us state the result thus obtained as a Lemma.

{\bf Lemma.} {\it The eigenvalues of operators $ \varepsilon P_0 A'$ and $-i/(\gamma \hbar)\hat H$ coincide.
The operator $\varepsilon P_0 A'$ has a complete system of eigenfunctions on the subspace of values of the projector $P_0$.
}

Consider the eigenvalue problem for equation~(\ref{eq_diff'}) of the form
 $\lambda_\varepsilon \varphi_\varepsilon=(\varepsilon A'+B')\varphi_\varepsilon.  $

According to perturbation theory, let us look for solutions as series:
$$
\lambda_\varepsilon=\lambda_0+\varepsilon\lambda_1+ \varepsilon^2\lambda_2+ ...
$$
$$
\varphi_\varepsilon=\varphi_0+\varepsilon\varphi_1+\varepsilon^2\varphi_2+...
$$
Let us substitute these series into the eigenvalue equation and compare coefficients before equal powers
of $\varepsilon.$  We obtain:
\begin{eqnarray}
 (\varepsilon_0)&&\ \ \ \ \ \ \  \lambda_0 \varphi_0=B' \varphi_0;\nonumber\\
 (\varepsilon_1)&&\ \ \ \ \ \ \  \lambda_1 \varphi_0+\lambda_0 \varphi_1=A' \varphi_0+B' \varphi_1;\nonumber
\end{eqnarray}

We are interested in perturbations of eigenvalue $  \lambda_0 =0$. In this case equation~$(\varepsilon_0)$
means that $\varphi_0$ is an eigenfunction of operator $B'$ with eigenvalue~$0$.

Let us apply the operator $\varepsilon P_0$ to both parts of equality~$(\varepsilon_1)$.
Taking into account the equalities $P_0\varphi_0=\varphi_0$,
$  \lambda_0 =0$, and $P_0 B'=0$, we obtain:
$$  \varepsilon \lambda_1 \varphi_0= \varepsilon P_0 A' \varphi_0.
$$

Hence the quantity $  \varepsilon \lambda_1$ is an eigenvalue of the operator $\varepsilon P_0 A'$ on the subspace of
values of the projector $P_0$, and by the Lemma the quantity $  \varepsilon \lambda_1$
is an eigenvalue of the operator
$-i/(\gamma \hbar)\hat H$. On the other hand, the quantity  $  \varepsilon \lambda_1$ is by definition a first
order correction to the eigenvalue 0 of the operator
$\varepsilon A'+B'$.

Thus, the first part of Theorem~4 is proved.

To prove the second part, let us present an arbitrary function $\varphi_0(t')$ from the subspace of values of the
projector $P_0$
as a sum (or integral for continuous spectrum):
$$
\varphi_0(t') =\sum_{k=1}^\infty  c_k(t') \varphi_0^k,
$$
where $\varphi_0^k$ are the eigenfunctions of the operator $ P_0 A'$ with eigenvalues $\lambda_1^k$.
This is possible, since by the Lemma such functions form a complete system. Besides that, the same Lemma implies that
the numbers $\lambda_1^k$ are pure imaginary, since they are eigenvalues of the operator
$-i/(\varepsilon \gamma \hbar)\hat H$, where $\hat H$ is a self adjoint Hamilton operator with real
spectrum.

Let $\lambda^k_\varepsilon\approx\varepsilon\lambda^k_1+ \varepsilon^2\lambda^k_2,\ \ $ and
$\ \ \varphi^k_\varepsilon\approx\varphi^k_0+\varepsilon\varphi^k_1$
be the  approximations of eigenvalues and eigenfunctions corresponding to eigenfunctions $\varphi_0^k$.
By assumptions of Theorem~4 the real parts $Re \lambda^k_2$ are different for various $k$.

Consider equation~(\ref{eq_diff'}) restricted to the subspace spanned by the basis vectors of the form
$\varphi^k_\varepsilon$ with the function of the form
$$\varphi (t') =\sum_{k=1}^\infty  c_k(t') \varphi^k_\varepsilon.$$
In this basis the equation splits (up to terms of order~$\varepsilon^3 $) into a system of equations
numbered by $k$ of the form:
$$\frac {\partial c_k(t')}{\partial t'}=(\varepsilon\lambda^k_1+ \varepsilon^2\lambda^k_2)c_k(t'),$$
whose solution reads as follows: $c_k(t')=c_k(0)\exp( (\varepsilon\lambda^k_1+ \varepsilon^2\lambda^k_2)t')$.
Hence, the solution of equation~(\ref{eq_diff'}) is approximately presented in the following form:
\begin{eqnarray}\label{varphi_t'}
\varphi(t')\!\!\!\!&=&\!\!\!\!\sum_{k=1}^\infty  c_k(0) \exp( (\varepsilon\lambda^k_1+ \varepsilon^2\lambda^k_2)t')
\varphi^k_\varepsilon \nonumber\\
\!\!\!\!&=&\!\!\!\!\sum_{k=1}^\infty  c_k(0) \exp( (\varepsilon\lambda^k_1+ \varepsilon^2\lambda^k_2)t')
(\varphi^k_0+\varepsilon\varphi^k_1)\nonumber\\
\!\!\!\!&=&\!\!\!\!\sum_{k=1}^\infty  c_k(0) \exp( (\varepsilon\lambda^k_1+ \varepsilon^2\lambda^k_2)t')\varphi^k_0+
    \sum_{k=1}^\infty  c_k(0) \exp( (\varepsilon\lambda^k_1+ \varepsilon^2\lambda^k_2)t')\varepsilon\varphi^k_1\nonumber\\
\!\!\!\!&=&\!\!\!\!\varphi_0(t')+\varepsilon\varphi_1(t'),
\end{eqnarray}
\begin{eqnarray}\label{varphi_0t'}
\mbox{where}\ \ \ \ \ \ \varphi_0(t')=\sum_{k=1}^\infty  c_k(0) \exp( (\varepsilon\lambda^k_1+ \varepsilon^2\lambda^k_2)t')
\varphi^k_0;\\
\varphi_1(t')= \sum_{k=1}^\infty  c_k(0) \exp( (\varepsilon\lambda^k_1+ \varepsilon^2\lambda^k_2)t')\varphi^k_1.
\end{eqnarray}

Since $\varepsilon $ is small, equality~(\ref{varphi_t'}) implies that the solution $\varphi(t')$
has a small difference with the function $\varphi_0(t')$ given by expression~(\ref{varphi_0t'}).

Since $\varepsilon\lambda^k_1$ is imaginary, each summand in the sum giving the function
$\varphi_0(t')$ in expression~(\ref{varphi_0t'}), decreases in time $t'=\gamma t$ proportionally to
$$
\exp(\varepsilon^2Re( \lambda^k_2)t')=\exp(\gamma \varepsilon^2Re( \lambda^k_2)t),
$$ where $Re( \lambda^k_2)$
is the real part of the number $\lambda^k_2$. This implies that after the time
$t\sim 1/(\gamma \varepsilon^2)$ this sum will be determined by the summand with the maximal number $Re( \lambda^k_2)$
among the summands with the valuable coefficients $ c_k(0).$

Thus, Theorem~4 is proved.


\begin{thebibliography}{99}
\bibitem{ben_dan}Beniaminov E. M.
{\it Diffusion processes in phase spaces and quantum mechanics}
 // Doklady Mathematics (Proceedings of the Russian Academy of Sciences), 2007, vol.76, No. 2, 771--774.
 \bibitem{ben_expand}Beniaminov E.~M.
{\it Quantization as asymptotics of a diffusion process in phase space}//
http://beniaminov.rsuh.ru/ExpandedDAN.pdf (in Russian; English translation: http://arXiv.org/abs/0812.5116v1). (2008)
\bibitem{feinman} Feynman R., Hibbs A. {Quantum mechanics and path integrals}, New York: McGraw-Hill, 1965.
\bibitem{kramers}Kramers H.A. // Physica. 1940. Vol. 7. P. 284-304.
\bibitem{vankampen}
{Van Kampen N.G.} {Stochastic Processes in Physics and Chemistry.} North Holland, Amsterdam, 1981.
\bibitem {kamke}
Kamke E., Differentialgleichungen: Losungsmethoden und Losungen, I, Gewohnliche Differentialgleichungen, B. G. Teubner, Leipzig, 1977. 
\bibitem{wigner}
{Wigner~E.}
{\it On the Quantum Correction For Thermodynamic Equilibrium}
 // Phys. Rev. 1932. V. 40. P. 749-759.
\bibitem{ben_arxiv}
{Beniaminov~E.M.}
{\it A Method for  Justification of the View of Observables in Quantum Mechanics and Probability Distributions in
Phase Space}
http://arxiv.org/abs/quant-ph/0106112  (2001).
\bibitem{zeh}
Zeh H.D. {\it Roots and Fruits of Decoherence.} // In:
Quantum Decoherence, Duplantier, B., Raimond, J.-M., and Rivasseau, V., edts. (Birkh\"auser, 2006), p. 151-175 (arXiv:quant-ph/0512078v2).
\bibitem{zurek}
Zurek W. H. {\it Decoherence and the transition from quantum to classical - REVISITED}
 arXiv:quant-ph/0306072v1.   2003 (An updated version of PHYSICS TODAY, 44:36-44 (1991)).
\bibitem{mensky}
Menskij M. B. {\it Dissipation and decoherence of quantum systems.}  Physics-Uspekhi (Advances in Physical Sciences), 2003,
vol.173, 1199-1219.
\bibitem{Shlyajx}
Shlyajh V. P. Quantum optics in the phase space, Fizmatlit, Moscow, 2005. 760 pp. (in Russian).
\bibitem{manko}
 Ibort A., Man'ko V.I., Marmo G., Simoni A., Ventriglia F.
  {\it On the Tomographic Picture of Quantum Mechanics.}  Phys.Lett.A374:2614-2617, 2010. arXiv:1004.0102v1.
\bibitem{khrennikov}
Khrennikov A. Quantum Randomness as a Result of Random
Fluctuations at the Planck Time Scale?  Int. J. Theor. Phys. 2008. {\bf 47}, N 1, P.114-124.
  arXiv:hep-th/0604011v3.
\bibitem{maslov1}
{ Maslov V. P.}
{\it Kolmogorov--Feller equations and the probabilistic model of quantum mechanics.}
 // Itogi nauki i tehniki. Probability, Mathematical Statistics and Cybernetics, 1982, vol. 19, p.~55--85 (in Russian).
\bibitem{maslov2}
{Maslov V. P.} {Quantization of thermodynamics and ultrasecondary quantization,}
Moscow: Institute for Computer Studies, 2001, 384 pp. (in Russian).
\end{thebibliography}
\end{document}